\documentclass[12pt]{article}

\usepackage{amsmath,amssymb,amsbsy,amsfonts,amsthm,latexsym,
                    amsopn,amstext,amsxtra,euscript,amscd}

\newtheorem{theorem}{Theorem}
\newtheorem{lemma}[theorem]{Lemma}

\newtheorem{corollary}[theorem]{Corollary}

\newtheorem{definition}[theorem]{Definition}

\def\cA{{\mathcal A}}
\def\cB{{\mathcal B}}
\def\cE{{\mathcal E}}
\def\cF{{\mathcal F}}
\def\cG{{\mathcal G}}
\def\cL{{\mathcal L}}
\def\cP{{\mathcal P}}

\def\cU{{\mathcal U}}
\def\cX{{\mathcal X}}

\newcommand{\remove}[1]{}

\newcommand{\tr}{{\mathrm{tr}}}
\newcommand{\Tr}{{\mathrm{Tr}_{\F_q/\F_p}}}

\def\({\left(}
\def\){\right)}
\def\<{\langle}
\def\>{\rangle}
\def\vec#1{\mathbf{#1}}

\def\rf#1{\left\lceil#1\right\rceil}

\def \F{\mathbb{F}}

\def \N{\mathbb{N}}

\def \R{\mathbb{R}}
\def \C{\mathbb{C}}

\def\ep{{\mathbf{\,e}}_p}
\def\en{{\mathbf{\,e}}_n}

\def\e0{\mathbf{\,e}}

\newcommand{\ket}[1]{\left| #1 \right \rangle}
\newcommand{\bra}[1]{\left\langle #1 \right|}
\newcommand{\sket}[1]{| #1 \rangle}
\newcommand{\sbra}[1]{\langle #1 |}

\def\vb{\psi}
\newcommand{\proj}[1]{\ket{#1}\bra{#1}}
\newcommand{\otop}[2]{\genfrac{}{}{0pt}{}{#1}{#2}}

\begin{document}
         \title{On Approximately Symmetric Informationally Complete
           Positive Operator-Valued Measures and Related Systems of Quantum
           States}


              \author{{\sc Andreas Klappenecker}\\
Department of Computer Science\\
Texas A\&M University\\
College Station, TX, 77843--3112, USA\\
\texttt{klappi@cs.tamu.edu}
\and
{\sc Martin R{\"o}tteler}\\
            NEC Laboratories America, Inc.\\
Princeton, NJ 08540, U.S.A.\\
            {\tt mroetteler@nec-labs.com}
\and
{\sc Igor E.~Shparlinski}\\
Department of Computing\\
            Macquarie University,\\ Sydney, NSW 2109, Australia\\
            {\tt igor@ics.mq.edu.au}
\and
{\sc Arne Winterhof}\\
Johann Radon Institute for\\ Computational
and Applied Mathematics \\
Austrian Academy of Sciences\\ Altenbergerstr.\ 69, 4040
Linz, Austria\\
{\tt arne.winterhof@oeaw.ac.at}
}
\maketitle

\begin{abstract}
          We address the problem of constructing positive operator-valued
          measures (POVMs) in finite dimension $n$ consisting of $n^2$
          operators of rank one which have an inner product close to uniform.
          This is motivated by the related question of constructing symmetric
          informationally complete POVMs (SIC-POVMs) for which the inner
          products are perfectly uniform. However, SIC-POVMs are
          notoriously hard to construct and despite some success of
          constructing them numerically, there is no analytic construction
          known. We present two constructions of approximate versions of
          SIC-POVMs, where a small deviation from uniformity of the inner
          products is allowed. The first construction is based on selecting
          vectors from a maximal collection of mutually unbiased bases
          and works whenever the dimension of the system is a prime power.
          The second construction is based on perturbing the
          matrix elements of a subset of mutually unbiased bases.

	Moreover, we construct vector systems in $\C^n$
          which are almost orthogonal and which might turn out to be useful
          for quantum computation.
          Our constructions are based on results of
          analytic number theory.
\end{abstract}

\section{Introduction}

\subsection{Background}

A basic question in quantum mechanics is how to obtain information
about the state of a given physical system by using suitable
measurements. Even in case many identically prepared copies of the
system are available, it is a nontrivial task to devise a measurement
procedure which uniquely identifies the given quantum state from the
statistical data produced by the measurements.  Note that this holds
true even in case the complete statistics, that is, the probabilities
for the different measurement outcomes, is known.

We next describe the possible measurements of the quantum system in
more detail and first remark that all systems considered in this paper
are of finite dimension~$n$.  If the state of the quantum system is
given by an $n\times n$ density matrix, then the complete measurement
statistics of one fixed {\em von Neumann measurement} is not sufficient to
reconstruct the state.  Indeed, this follows from the fact that the
statistics of a fixed von Neumann measurement determines at most
$n-1$ real parameters (specified by the probabilities of the
measurement outcomes), whereas a general density matrix is determined
by $n^2-1$ free real parameters.

In fact it is possible to perform a more general measurement procedure
on a quantum system, namely a {\em positive operator-valued measure},
or POVM for short, see~\cite{Nielsen,Peres}. A POVM is described by a
collection of positive operators $E_i\ge 0$, called POVM elements,
that partition the identity, that is, $\sum_i E_i = I_n$.  If the state
of the quantum system is given by the density matrix $\rho$, then the
probability $p_i$ to observe outcome $i$ in the POVM is given by the
Born rule
\begin{equation}\label{born}
p_i = \tr(\rho E_i),
\end{equation}
where $\tr(A)$ denotes the trace of a complex matrix $A$.
The task is to devise a POVM with operators $E_i$ such that the state
$\rho$ is uniquely specified by the probabilities $p_i$
in~\eqref{born}. The POVM is then called {\em informationally
   complete}, or simply an IC-POVM, and they appear to have been first
studied in~\cite{Prugovecki}. A particularly interesting question is
whether a POVM exists on $\C^n$ that consists of $n^2$ POVM elements
$E_i$ of rank one. Counting the number of parameters determined by the
measurement, we see that $n^2$ is indeed the minimal possible number
of such POVM elements.  In this case $E_i$ is a subnormalized
projector, that is, $E_i = \Pi_i/n$ for projectors $\Pi_i =
\ket{\psi_i}\bra{\psi_i}$ corresponding to some vectors $\ket{\psi_i}$
in $\C^n$.  In~\cite{CFS} it has been shown that IC-POVMs exist in all
dimensions and in~\cite{ArPeSa} a method has been given how to
construct IC-POVMs by taking a fixed fiducial start vector
$\ket{\psi}$ and taking the orbit of this vector under a (projective)
group operation.

As an example of this type, consider the normalized states
$\ket{\psi_1}, \ldots, \ket{\psi_4}$ defined as follows:
\begin{align*}
\ket{\psi_1} & = \displaystyle{\frac{1}{3}} \left( \begin{array}{c} 1 \\ 2{+}2i
          \end{array} \right), \quad
&\ket{\psi_2}  = \frac{1}{3} \left( \begin{array}{c} 2{+}2i \\ 1
          \end{array} \right), \\
         \ket{\psi_3} & = \displaystyle{\frac{1}{3}} \left( 
\begin{array}{c} 1 \\
{-}2{-}2i
          \end{array} \right), \quad
&\ket{\psi_4}  = \frac{1}{3}\left( \begin{array}{c} 2{+}2i \\
{-}1
          \end{array} \right).
\end{align*}
Then the rank one operators defined by $E_i = 1/2
\ket{\psi_i}\bra{\psi_i}$ are given by
\begin{align*}
         E_1 & = \displaystyle{\frac{1}{18}} \left( \begin{array}{cc} 
1 & 2+2i \\
2-2i & 8 \end{array} \right),   \qquad  \
           E_2 = \frac{1}{18} \left(
\begin{array}{cc} 8 & 2-2i \\ 2+2i & 1 \end{array} \right),\\
E_3 & =  \displaystyle{\frac{1}{18}} \left( \begin{array}{cc} 1 &
-2-2i \\ -2+2i & 8 \end{array} \right),   \quad
         E_4   = \frac{1}{18}\left( \begin{array}{cc} 8 & -2+2i \\
-2-2i & 1 \end{array} \right),
\end{align*}
and it can be verified easily that $E_1+E_2+E_3+E_4=I_4$ is the
identity matrix and that the matrices $E_1,E_2,E_3, E_4$ are linearly
independent. For the possible inner products between two POVM elements
$E_i$ and $E_j$ where $i \ne j$ we obtain that ${\tr}(E_i E_j) \in
\{4/81, 49/324\}$.

Our goal is to find IC-POVMs on $\C^n$ such that
$n^2\tr(E_iE_j)$
is ``small'' for distinct POVM elements
$E_i=|\psi_i\rangle\langle \psi_i|/n$
and
$E_j=|\psi_j\rangle\langle \psi_j|/n$.
In Section~\ref{previous} we make precise what we mean by the inner
products being small and briefly summarize previous work on the
problem.

\subsection{Notation}

         We use the Landau notation to compare the
asymptotics of two functions $f, g: \N \rightarrow \C$.
We recall that
$f(n)=o(g(n))$ means
$\lim_{n \to \infty}  f(n)/g(n) = 0$.
         Furthermore,
$f(n)=O(g(n))$ means that there
exists a constant $c > 0$,
such that $|f(n)| \leq c g(n)$ for all
$n \ge 1$.
Throughout the paper, the implied constants in the symbols
`$o$' and `$O$' may occasionally, where obvious, depend on an integer
parameter $d$ and a small real parameter $\varepsilon > 0$, and are
absolute otherwise.

For an integer $n \geq 1$ we denote by $\C^n$ the $n$-dimensional
vector space over the complex numbers $\C$. For two vectors $
\ket{\psi} = (a_1, \ldots, a_n)\in \C^n$ and $\ket{\varphi} = (b_1,
\ldots, b_n) \in \C^n $, we use
$$
\langle \psi | \varphi \rangle = \sum_{i=1}^n \overline{a}_i\, b_i
$$
to denote their inner product. We also define $ \delta_{i,j}= 1$ if
$i=j$ and $0$ otherwise.  We denote the identity matrix of size $n
\times n$ by $I_n$, the all-ones matrix of size $n$ by $J_n = [ 1
]_{i,j=1}^n$.  We use ${\mathrm{diag}}(a_1, \ldots, a_n)$ to denote
the $n \times n$ diagonal matrix which has $a_1, \ldots, a_n$ on the
main diagonal. For matrices $A$ and $B$, we use $A \oplus B$ to
denote their block-diagonal direct sum.

For a real $z$ and an integer $m$ we use the notation $
{\mathbf{\,e}}_m(z) = \exp(2 \pi \iota z/m), $ where $\iota =
\sqrt{-1}$.

We use $\F_q$ to denote the finite field of $q$ elements, and we also
assume that for a prime $p$, the field $\F_p$ is represented by the
set $\{0, \ldots, p-1\}$.

As we have mentioned, we use $\tr(A)$ to denote the trace of a
complex matrix $A$. On the other hand, for an element $a \in \F_q$
we use $\Tr(a)$ to denote its trace in the prime subfield $\F_p$ of $\F_q$,
see~\cite{LN}.
That is, if $q = p^m$ for a prime $m$, then
$$
\Tr(a) =\sum_{j=0}^{m-1} a^{p^j}.
$$

\subsection{Previously Known Results}
\label{previous}

A particularly appealing case of IC-POVMs arises
when furthermore all of the inner products of the vectors $\ket{\psi_i}$
are small. An extremal case in this sense arises when we are given a system
of $n^2$ normalized
vectors $\{ \ket{\psi_i}  \ : \  i = 1, \ldots, n^2 \}$ in $\C^n$
for which
\begin{equation}\label{sicpovm}
|\langle \psi_i | \psi_j \rangle|^2 = \frac{1}{n+1},
\qquad
1\le i<j\le n^2.
\end{equation}
Indeed, for any system of $k$ vectors $\ket{\psi_1}, \ldots,
\ket{\psi_k}$ in $\C^n$ for which the
absolute values of
pairwise inner products are
constant $|\langle \psi_i | \psi_j \rangle|^2 = \alpha$, where $\alpha
\in \R$ (for $1 \leq i < j \leq k$) the so-called special bound
holds~\cite{Hoggar2} which says that $k \leq \frac{n(1-\alpha)}{1-n
\alpha}$. Specializing $\alpha = 1/(n+1)$ we obtain that $n^2$ is the
largest possible number of vectors satisfying~\eqref{sicpovm}.

A system of vectors as in~\eqref{sicpovm}, respectively the
corresponding POVMs, are called {\em symmetric informationally
   complete POVMs}, or SIC-POVMs for short. They have several very
desirable properties, see~\cite{CFS} for a discussion in the context
of the quantum de~Finetti theorem and more generally in their Bayesian
approach to quantum mechanics and its interpretation~\cite{CFSBayes}.
Furthermore, see~\cite{Fuchs,FuSa} for their role in establishing the
quantumness of a Hilbert space and the related question about optimal
intercept-resend eavesdropping attacks on quantum cryptographic
schemes. Explicit analytical constructions of sets
satisfying~\eqref{sicpovm} have been given for dimensions $n=2,3,4,5$
in~\cite[Section 3.4]{Zauner:99} and~\cite{RBSC}, for dimension $n=6$
in~\cite{Gra}, for dimensions $n=7,19$ in~\cite{Appleby} and for
dimension $n=8$ see~\cite{Hoggar}.  While it has been conjectured that
SIC-POVMs exist in all dimensions~\cite[Section 3.4]{Zauner:99}
or~\cite{RBSC} and numerical evidence exists for dimensions up to
$45$, see~\cite{RBSC}, it is a difficult task to explicitly construct
systems of vectors which satisfy~\eqref{sicpovm}.  There are no known
infinite families of SIC-POVMs and in fact it is not even clear if
there are SIC-POVMs for infinitely many $n$.

\subsection{Our Results}

In the first part (Section~\ref{asics}) of this paper we relax
condition~\eqref{sicpovm} on the inner products slightly and allow that
\begin{equation}\label{asicpovm}
          |\langle \psi_i | \psi_j \rangle|^2
\le \frac{1+o(1)}{n}, \qquad
1\le i<j\le n^2.
\end{equation}

The purpose of the first part of this paper is to show that infinite
families of systems of $n^2$ normalized vectors which
satisfy~\eqref{asicpovm} and give rise to IC-POVMs exist. We call the
rank one projectors obtained from such systems of vectors {\it
   approximately symmetric informationally complete POVMs\/}, or
ASIC-POVMs for short.  Here we show that when $n=p^r$ is a power of a
prime $p$, ASIC-POVMs can be constructed.

In the second part (Section~\ref{relaxed}) of the paper we explore
properties of other approximately symmetric vector systems where we do
not require the properties of completeness and informational
completeness but require the property of approximate symmetry.  We
also relax~\eqref{asicpovm} further by allowing that the inner
products be bounded from above by $|\langle \psi_i | \psi_j \rangle|
\leq (2+o(1))/\sqrt{n}$ and by dropping the requirement that the
vectors give rise to a POVM. This additional freedom then allows us to
construct bases in all dimensions $n$.  Besides their general
mathematical interest in constructing such vector systems they might
be useful in quantum cryptographic scenarios which generalize the BB84
setting~\cite{bennett84}, such as the protocols described
in~\cite{BTBLR,Renes,RenesPRA}.  See also~\cite{CRE} for an analysis
of general schemes for quantum key distribution where the sender uses
arbitrary quantum states and the receiver's measurement is replaced by
a POVM.

Besides approximations to SIC-POVMs we also consider approximations to
{\em mutually unbiased bases} (MUBs). Since we also need MUBs for our
construction of ASIC-POVMs we briefly recall their definition.  A
maximal set of MUBs is given by a set of $n^2+n$ vectors in $\C^n$
which are the elements of $n+1$ orthonormal bases $ \cB_k =
\{\ket{\psi_{k,1}}, \ldots, \ket{\psi_{k,n}}\}$ of $\C^n$ where $k =0,
\ldots, n$. Hence,
\begin{equation}
\label{eq:delta}
\langle  \psi_{k,i}| \psi_{k,j} \rangle = \delta_{i,j},
\end{equation}
and the defining property is the mutual unbiasedness,
given by
\begin{equation}
\label{eq:type1}
|\langle  \psi_{k,i}| \psi_{\ell,j} \rangle| = \frac{1}{\sqrt{n}}
\end{equation}
for $0 \leq k ,\ell\leq n$, $k \neq \ell$, and $1 \leq i,j \leq n$.
Starting from~\cite{Ivanovic:81,schwinger60,WoFi:89} an extensive
growing body of research explores MUBs and their constructions,
see~\cite{AsChWo,BBRV:2001,BeEr,Chaturvedi:2002,Durt:2004,HKCSS:94,KlRo,PR:2003,PRPS,WoBe}
and references therein.  However, so far maximally sets of $n+1$ MUBs
in dimension $n$ are only known to exist in any dimension $n=p^r$
which is a power of a prime $p \ge 3$, see~\cite{KlRo} for an overview
and some of such constructions. The main construction is based on
Gaussian sums and in the case of prime $n = p$ can be described as
$$
\ket{\psi_{k,j}} = \frac{1}{\sqrt{p}}\(\ep(ku^2 + ju)\)_{u =1}^p,
\qquad 1\leq k,j \leq p,
$$
and also $\cB_{0}$ being a standard orthonormal basis, that is, $
\ket{\psi_{0,j}} = \(\delta_{j,u}\)_{u=1}^p$.

One can use additive characters over an arbitrary finite field to
extend this construction to an arbitrary prime power $n = p^r$.
However the condition that $n=p^r$ is a prime power is still somewhat
too restrictive and unnatural for quantum computation. So a natural
question arises whether MUBs exist for every positive integer $n$. In
the second part we consider vector systems where we relax the
conditions~\eqref{asicpovm} and~\eqref{eq:type1}. We use exponential
sums to construct vector systems for any dimension $n$ which

\begin{itemize}
\item satisfy~\eqref{eq:delta} but instead of~\eqref{eq:type1} all
          other inner products are $O(n^{-1/4})$;
\item is normalized but instead of~\eqref{sicpovm} all other inner
          products are at most $(2 + o(1)) n^{-1/2}$.
\end{itemize}

We call vector systems of $n^2+n$ vectors in $\C^n$ which satisfy
\eqref{eq:delta}, and instead of \eqref{eq:type1} the condition
$$
          |\langle \psi_i | \psi_j \rangle|^2
\le \frac{1+o(1)}{n},
\qquad
1\le i<j\le n^2+n
$$
{\it approximately mutually unbiased bases\/}, or AMUBs for short.

We also construct some vector systems using multiplicative and mixed
character sums, which
\begin{itemize}
\item satisfy~\eqref{eq:delta} and assuming some natural and widely
          believed conjecture on the distribution of primes in arithmetic
          progressions all other inner products are $O(n^{-1/2}\log n)$;
\item
in the special case of   $n = p-1$ where $p$ is a prime,
form AMUBs.
\end{itemize}

Interestingly, our arguments use both the classical bound of
Weyl~\cite{Weyl} (see also~\cite{Hua,Vau}) as well as the more recent,
but no less celebrated, bounds of Weil~\cite{Weil1} (see
also~\cite{Li,LN,Weil2}).  Besides exponential sum techniques we also
use some recent results about the gaps between prime numbers
from~\cite{BaHaPi}. We conclude with some conjectures and open
questions concerning our constructions in Section~\ref{remarks}.

\section{Constructing ASIC-POVMs}\label{asics}

\subsection{Preliminaries}

We begin by giving a definition of the vectors and associated rank one
operators we are interested in.

\begin{definition}[ASIC-POVMs]\label{def:asic-povm}
   Suppose that $n$ is a positive integer. Let $\cA = \{ \ket{\psi_i} \
   : \ i = 1, \ldots, n^2\}$ be a set of vectors in $\C^n$.  Let $\cE =
   \{ E_i = \ket{\psi_i}\bra{\psi_i}/n \ : \ i = 1, \ldots, n^2\}$ be
   the corresponding set of subnormalized projection operators.  If
   $\cE$ satisfies the conditions
\begin{itemize}
\item[$(i)$] $\sum_{i=1}^{n^2} E_i = I_n$
         ({\em completeness/POVM condition});
\item[$(ii)$] the matrices $E_i$ are
         linearly independent as elements of $\C^{n \times n}$
         ({\em informational completeness});
\item[$(iii)$]
        $|\langle \psi_i | \psi_j \rangle|^2 = n^2 {\rm tr}(E_i
E_j) \le (1 + o(1)) n^{-1}$
for
$1 \leq i < j \leq n^2$ ({\em approximate symmetry});
\end{itemize}
then we call $\cE$ an
{\em approximately symmetric informationally complete
POVM}, or ASIC-POVM for short.
\end{definition}

We remark, that in fact, sometimes we also
refer to the corresponding set $\cA$ as an ASIC-POVM.

In the subsequent sections, we present two different constructions
that give rise to infinite families of ASIC-POVMs. The first
construction is based on the observation that a set of $n+1$ mutually
unbiased bases in $\C^n$ gives rise to an IC-POVM,
cf.~\cite{Ivanovic:81,WoFi:89}. However, this IC-POVM consists of
$n^2+n$ rank-one operators; thus, it is an overcomplete generating set
of the vector space of all $n\times n$ matrices. In our first
construction in Section~\ref{ASICselect} we show how to select $n^2$
projectors that allow us to derive an ASIC-POVM. The second
construction in Section~\ref{ASICperturb} starts from all vectors
contained in $n$ of the $n+1$ MUBs. We show that by slightly
perturbing the components of these vectors it is possible to obtain an
ASIC-POVM.

\subsection{POVMs and Frames}\label{fuchsTrick}

Suppose that $\cA=\{\ket{\psi_i}\in \C^n \colon 1\le i\le n^2\}$
is a system of $n^2$ vectors of unit norm, such that
$\cA$ spans $\C^n$ and
the associated subnormalized projectors
$E_i=\ket{\psi_i}\bra{\psi_i}/n$ satisfy $n^2\tr(E_iE_j) =
(1+o(1))n^{-1}$ whenever $i\neq j$.

We would like to have that the subnormalized projectors $E_i$ form a
POVM, but, unfortunately, the completeness relation $\sum_{i=1}^{n^2}
E_i = I_n$ is in general not satisfied.  However, there is a way to
fix this using a technique described in~\cite{CFS}: Define a positive
semidefinite hermitian operator $G$ by
$$
G = \sum_{i=1}^{n^2} E_i.
$$
Since $\cA$ spans $\C^n$, the inequality $\langle \varphi | G |
\varphi \rangle = \sum_i |\langle \psi_i | \varphi \rangle |^2 > 0$
holds for any nonzero vector $\ket{\varphi}$
in $\C^n$, so $G$ is even
positive definite. It follows that $G^{-1}$ exists and is positive
definite, and we can form the uniquely determined positive definite
square-root $G^{-1/2}$. The $n^2$ rank-one operators
$$\cE = \{ F_i=  G^{-1/2} E_i G^{-1/2}\colon 1\le i\le n^2\}$$
form a POVM, since $\sum_i F_i = G^{-1/2} G G^{-1/2} = I_n$.

Clearly, if the operators $E_i$ are linearly independent, then
so are
the operators $F_i$. Therefore, the procedure preserves
information-completeness.

In general, if we switch from the rank-one operators $E_i$ to the
rank-one operators $F_i$, then ${\tr}(F_i F_j) \leq (1+o(1))/n^3$
might not hold for some $i\neq j$. However, if $G^{-1}$ is close to
the identity matrix, then approximate symmetry is preserved as well.

We now mention some connections between POVMs and
the theory of frames~\cite{BeFi,CaKo,Daubechies}.

\begin{definition}[Frames] A set $\cF=\{ \ket{\psi_i}  \colon
1\le i\le N \}$ of vectors in $\C^n$ is called a {\em frame\/} if there
exist real numbers $a$ and $b$, with $0 <a \leq b $, such that
\[
a \langle \varphi | \varphi \rangle \leq \sum_{i=1}^N
|\langle \varphi | \psi_i \rangle |^2 \leq
b \langle \varphi | \varphi \rangle
\]
holds for all $\ket{\varphi} \in \C^n$.

If $a=b$, then the frame is called a {\em tight frame\/}, and
if $a=b=1$ then the frame is called a {\em Parseval frame}.
\end{definition}

We can associate with the frame $\cF$ its frame operator
$G=\sum_{k=1}^N \ket{\psi_k}\bra{\psi_k}$.
If we are given a frame $\cF$ with frame operator $G$, then
$$\cG=\{ G^{-1/2}\ket{v} \colon \ket{v}\in \cF\}$$ is a Parseval
frame, see~\cite[Theorem~4.2]{casazza00}. The
projectors associated with $\cG$ form a POVM, since
$ \sum_{v\in \cF} G^{-1/2}\ket{v}\bra{v} G^{-1/2} =
G^{-1/2}GG^{-1/2}=I$
holds.

If we have a Parseval frame $\cG$ in $\C^n$ with $n^2$ elements such
that the associated projection operators are linearly independent and
the frame elements satisfy the approximate symmetry~\eqref{asicpovm},
then the projectors corresponding to the frame $\cG$ form an
ASIC-POVM.

\subsection{Construction~I: Pruning MUBs}
\label{ASICselect}

The first construction of ASIC-POVMs is based on the idea to select a
suitable collection of $n^2$ vectors from a set of $n^2 + n$ vectors
that form a maximal set of $n+1$ mutually unbiased bases of $\C^n$.
Our goal is to choose $n^2$ vectors such that the corresponding
projection operators are linearly independent.  We recall a known fact
that belongs to the folklore of mutually unbiased bases; it is
implicitly contained in~\cite{Ivanovic:81,WoFi:89}, and more
explicitly in~\cite{Hayashi}, and our proof is based on the latter.

\begin{lemma}\label{MUBindep}
Suppose that $\cB_a=\{ v_{a,b} \ : \  0\le b<n\}$, with $0\le a\le n$, are
$n+1$ mutually unbiased bases of\/ $\C^n$, and let
$$\cP=\{ \proj{v_{a,b}}  \ : \  0\le a\le n, 0\le b<n \}$$ denote the
associated set of $n^2+n$ projectors.  The $n^2$ projection operators
in $\cP^*=\{ \proj{v_{a,b}} \ : \  (a,b)=(0,0) \text{ or } b\neq
0\}\subset \cP$ are linearly independent.
\end{lemma}
\begin{proof}
First, suppose that a linear relation
\begin{equation}
\label{lin_dep}
\sum_{a=0}^n\sum_{b=0}^{n-1} \gamma_{a,b}\proj{v_{a,b}}=0
\end{equation}
holds
for some $\gamma_{a,b}\in \C$. We are going to show that this has some
rather strong consequences for the coefficients $\gamma_{a,b}$. If we apply
the projection operators from $\cP$ and take the trace, then we obtain
a linear system of equations $A\vec{g}=0$, where
$$
A = \left[\tr( \proj{v_{a,b}}\, \proj{v_{c,d}})\right]_{(a,b),(c,d)}
$$
and $\vec{g}=(\gamma_{a,b})$ is a column vector. The matrix $A$ is
block-circulant,
$$
A= \left(\begin{array}{lclll}
I_n & \frac{1}{n}J_n & \cdots & \frac{1}{n}J_n & \frac{1}{n}J_n \\
\frac{1}{n}J_n & I_n & \cdots & \frac{1}{n}J_n & \frac{1}{n}J_n \\
\ddots &\ddots &  & \ddots & \ddots\\
\frac{1}{n}J_n & \frac{1}{n}J_n & \cdots & \frac{1}{n}J_n & I_n
\end{array}
\right),
$$
with $n\times n$ identity matrices in the diagonal blocks, and multiples of
the $n\times n$ all-one matrix in the off-diagonal blocks.

If we subtract two equations in $A\vec{g}=0$ that belong to the same block
row, then we find that $\gamma_{a,b}=\gamma_{a,d}$ holds for all $0\le b,d<n$
and all indices~$a$. Therefore, the coefficients $\gamma_{a,b}$ do not
depend on the value of $b$.

Finally, suppose that the left hand side of~\eqref{lin_dep} consists
of a linear
combination of projectors belonging to the set~$\cP^*$, meaning that
the coefficients
$\gamma_{a,b}=0$ when $a\neq 0$ and $b=0$. It follows that $\gamma_{a,b}=0$
holds whenever $a\neq 0$, since $\gamma_{a,b}=\gamma_{a,0}$ by our previous
observation. Therefore, the left hand side of~\eqref{lin_dep}  reduces to
$$
\sum_{b=0}^{n-1} \gamma_{0,b} \proj{v_{0,b}} = \sum_{b=0}^{n-1}
\gamma_{0,0} \proj{v_{0,b}} = 0.$$
Thus, we must have
$\gamma_{0,0}=0$. Therefore, we can conclude that the projectors in $P^*$
are linearly independent, as claimed.
\end{proof}

We also recall the basic construction of
MUBs in prime power dimension; see, for instance,~\cite{KlRo,WoFi:89}.

\begin{lemma}
\label{MUBs}
Let $q$ be a power of a prime $p\ge 3$. Define
\[
         \ket{\psi_{a,b}} = q^{-1/2} \(\ep\(\Tr(ax^2+bx)\)\)_{x\in \F_q}\in
\C^q. \]
Then the standard basis $\cB_0$ together with the bases $\cB_a=\{
\ket{\psi_{a,b}} \ : \ b\in
\F_q\}$, $a\in \F_q$, form a set of $q+1$ mutually unbiased bases of $\C^{q}$.
\end{lemma}

Our first construction of ASIC-POVMs is given in the next theorem.

\begin{theorem}\label{constructionI}
Let $q$ be a power of a prime $p\ge 3$. Let
$$\ket{\psi_{a,b}} = q^{-1/2} \(\ep\(\Tr(ax^2+bx)\)\)_{x\in \F_q}\in \C^q
$$
for
all $(a,b)\in \F_q\times \F_q^\times$ and
$\ket{\psi_{a,0}} = (\delta_{a,x})_{x \in \F_q}$ for all $a\in \F_q$.
We define $E_{a,b}
= \ket{\psi_{a,b}}\bra{\psi_{a,b}}/q$ and
$$G
= \sum_{a\in \F_q} \sum_{b\in \F_q}  E_{a,b}.
$$ Then the set $\{ F_{a,b} \ : \ a,b\in \F_q\}$, with $F_{a,b} =
G^{-1/2} E_{a,b} G^{-1/2}$, is an ASIC-POVM.
\end{theorem}

\begin{proof}
The linear independence of the operators $E_{a,b}$ follows from
Lemma~\ref{MUBindep}. It remains to show that the matrix $G^{-1}$ is
close to
the identity and that $F_{a,b} =
G^{-1/2}\ket{\psi_{a,b}}\bra{\psi_{a,b}}G^{-1/2}$ indeed forms an
ASIC-POVM. First, we explicitly compute the frame operator $G$.
We have
\begin{eqnarray*}
G & = &
\frac{1}{q}\sum_{(a,b)\in \F_q\times \F_q^\times}
|\psi_{a,b}\rangle\langle \psi_{a,b}|\; + \frac{1}{q}I_q \\
          &=  & \frac{1}{q^2} \Bigg( \sum_{x,y,a,b\in \F_q}
            \ep\(\Tr(a(x^2{-}y^2) + b(x{-}y))\) \ket{x}\bra{y}\\
           && \qquad- \sum_{x,y,a\in \F_q} \ep\(\Tr(a(x^2{-}y^2))\)
\ket{x}\bra{y}\Bigg)
              + \frac{1}{q} I_q.
\end{eqnarray*}
We notice that $\langle x\,|\,G|\,x\rangle=1$ for $x\in \F_q$,
$\bra{x}G\ket{-x}=-1/q$ for $x\in \F_q^\times$, and $\langle x| G
|y\rangle=0$
for $x,y\in \F_q$ with $x\ne \pm y$.
Therefore, we can express the operator $G$ in
the form
$$G = I_q - \frac{1}{q} Q + \frac{1}{q} \ket{0}\bra{0},\quad
\text{where} \quad Q = \sum_{x\in \F_q} \ket{x}\bra{-x}.$$
Using the structure of $G$, it follows that the inverse is
given by
$$
G^{-1} = \left(1+\frac{1}{q^2-1}\right) I_q +
\frac{q}{q^2-1} Q - \frac{1}{q-1} \ket{0}\bra{0}.
$$
Observe that the set of vectors
$\{\ket{\psi_{a,b}} \ : \ (a,b)\in \F_q\times \F_q^\times\}$
is invariant under~$Q$, since
        $Q \ket{\psi_{a,b}} = \ket{\psi_{a,-b}}$.
Recall that by Lemma~\ref{MUBs} the bounds
$$|\langle \psi_{a,b} | \psi_{c, d} \rangle|
        \leq q^{-1/2}$$
hold, whenever $(a,b) \ne (c,d)$.
Defining
$|\widetilde{\psi}_{a,b}\rangle = G^{-1/2} \ket{\psi_{a,b}}$
we get
$$F_{a,b}
= G^{-1/2} E_{a,b} G^{-1/2} = | \widetilde{\psi}_{a,b} \rangle
\langle\widetilde{\psi}_{a,b} |/q$$
and
obtain that
\begin{eqnarray*}
         q^2 \tr(F_{a,b} F_{c,d}) &=&  |\langle \widetilde{\psi}_{a,b} |
          \widetilde{\psi}_{c,d} \rangle|^2 = |\langle \psi_{a,b} |
          G^{-1} | \psi_{c,d} \rangle|^2 \\
          & \leq & \left|\left(1+\frac{1}{q^2-1}\right) \langle
\psi_{a,b} | \psi_{c,d} \rangle\right|^2 \\
& & +
          \left|\left(\frac{q}{q^2-1}\right)\langle \psi_{a,b} |
\psi_{c,-d} \rangle \right|^2 + \left(\frac{1}{q(q-1)}\right)^2 \\
          & \leq & \frac{1}{q} \left(1+\frac{1}{q^2-1}\right)^2 + \frac{1}{q}
          \left( \frac{q}{q^2-1} \right)^2 + \left(\frac{1}{q(q-1)}\right)^2 \\
& = & \frac{1 + o(1)}{q}.
\end{eqnarray*}
This shows that the rank-one operators $F_{a,b}$ form an ASIC-POVM.
\end{proof}

\subsection{Construction II: Perturbing MUBs}
\label{ASICperturb}

We now describe a second, different, method to obtain a set of $n^2$
vectors such that the corresponding projectors span the space of all
$n\times n$ matrices. This construction of ASIC-POVMs works for all
dimensions $n$ such that $n$ is an odd prime number.

We note that all arithmetic operations in any expression
involving elements of $\F_p$ and real numbers
are performed over the real numbers (where each element of
$\F_p$ is represented by an integer in the range $[0, p-1]$).
   For example, for a real $r\in \R$ and $a,x \in \F_p$, the power $r^{ax}$
means $r^u$, where the integer $u = ax$ can be of size $(p-1)^2$.

\begin{theorem}\label{constructionII}
          Let $p$ be
an odd
prime number, and let $0 < r < 1$ be a real number. For
          $a, b \in \F_p$ define
$$
\ket{\varphi_{a,b}} =
          \sqrt{\frac{1-r^{2a}}{1-r^{2pa}}}\, (r^{ax} \ep(ax^2+bx))_{x\in
          \F_p}\in \C^p.
$$
and let $E_{a,b} = \ket{\varphi_{a,b}}\bra{\varphi_{a,b}}/p$. Then the
$E_{a,b}$ are linearly independent. Furthermore, let
\[
G  = \sum_{a,b \in \F_p} E_{a,b}.
\]
Then for $r = 1 - p^{-3}$ the set $\{ F_{a,b} \ : \ a,b\in
\F_p\}$, with $F_{a,b} = G^{-1/2} E_{a,b} G^{-1/2}$ is an ASIC-POVM.
\end{theorem}

\begin{proof}
First, we show that the matrices $E_{a,b}$ are linearly independent.
Note that instead of considering the normalized vectors
$\ket{\varphi_{a,b}}$ it is possible to consider the vectors
$| \widetilde\varphi_{a,b}\rangle = (r^{ax} \ep(ax^2+bx))_{x\in \F_p}$ and
to show that the corresponding projectors $\widetilde E_{a,b} =
|\widetilde\varphi_{a,b}\rangle \langle \widetilde\varphi_{a,b} |$ are
linearly independent.

We use
the technique
       introduced in~\cite{ArPeSa} to check whether the
matrices $\widetilde E_{a,b}$ are linearly independent.  To each $n
\times n$ matrix $\widetilde E_{a,b}$ we associate a state
$|\widetilde E_{a,b}\rangle$ which is simply the row-wise
concatenation of the entries of $\widetilde E_{a,b}$ as a vector of
length $n^2$. Then the matrices $\widetilde E_{a,b}$ are linearly
independent if and
only if the matrix 
$S=\frac{1}{p}\sum_{a,b\in\F_p} \sket{\widetilde E_{a,b}}
\sbra{\widetilde E_{a,b}}$ has full rank. We obtain
        \begin{equation*}
\begin{split}
pS=& \sum_{a,b\in \F_p} \sket{\widetilde E_{a,b}} \sbra{\widetilde E_{a,b}} \\
         = &  \sum_{\otop{a\in \F_p}{x,y,u,v\in \F_p}}
r^{a(x+y+u+v)} \ep(a(x^2{-}y^2{-}u^2{+}v^2))\\
& \qquad \qquad \qquad \qquad
\times \sum_{b\in \F_p} \ep(b(x{-}y{-}u{+}v)) \ket{x,y} \bra{u,v} \\
         = & p 
	 \sum_{\otop{a\in \F_p}{x,y,u,v\in \F_p}} r^{a(x+y+u+v)}
\ep(a(x^2-y^2-u^2+v^2))
\delta_{x-y, u-v} \ket{x,y} \bra{u,v}.
\end{split}
\end{equation*}

The rows of $S$ are labeled by pairs $(x,y)$, with $x,y\in \F_p$, and
the columns by pairs $(u,v)$, with $u,v \in \F_p$. We first note that $S$
can be written as a block-diagonal matrix of $p$ submatrices, each of
size $p \times p$, if the rows and columns of $S$ are suitably
rearranged. For $0 \leq i \leq (p-1)$, we define the sets
$\cL_i = \{ (x, x+i)  \ : \  x \in \F_p \}$
(we treat $i$ as an element of $\F_p$ so $x+i$ is computed
in $\F_p$ too). We now order the rows and columns according to the list
$\cL = \bigcup_{i=0}^{p-1} \cL_i$. With respect to
this basis we obtain that
\[
S = A_0 \oplus A_1 \oplus \ldots \oplus A_{p-1},
\]
with $p\times p$ matrices
$$A_i = \left[ \sum_{a\in
\F_p} r^{a(x+y+u+v)} \ep(a(x^2{-}y^2{-}u^2{+}v^2)) \right]_{x,u \in \F_p}
$$
where $y=x+i$ and $v=u+i$. Hence, we obtain that
\[
A_i = \left[ \sum_{a\in \F_p} r^{2a(x+u+i)} \ep(2ai(u-x))
\right]_{x,u \in \F_p}
\]
and have to show that this matrix is invertible for $0 \leq i \leq
(p-1)$. In order to do so, we first observe that $x \mapsto x-i/2$
defines a permutation of the rows of any $p \times p$ matrix and that
similarly $u \mapsto u-i/2$ defines a permutation of the
columns (hereafter $i/2$ is computed in $\F_p$). Applying both the row
and the column permutation
to
$A_i$ we obtain the matrix
$$
B_i = \left[ \sum_{a\in \F_p} r^{2a(x+y)}
\ep(2ai(u-x))\right]_{x,u \in \F_p}.
$$
        Note further that $x \mapsto x/2$ and $y
\mapsto y/2$ induce permutations of the rows and columns of any $p
\times p$ matrix. Applying this to $B_i$ we obtain the matrix
\begin{eqnarray*}
          C_i
         &  =  &\left[\sum_{a \in \F_p} r^{a(u+x)}
\ep(ai(u-x))\right]_{x,u \in \F_p}\\
& = & \left[\sum_{a \in \F_p} (r^x \ep(-ix))^{a} (r^u \ep(iu))^{a}
              \right]_{x,u \in \F_p} \\
& = & \left[ (r \ep(-i))^{x k}\right]_{x, k=0}^{p-1}
\times \left[(
r \ep(i))^{\ell u}\right]_{\ell,u=0}^{p-1}.
\end{eqnarray*}

Since $|r \ep(i)| = |r \ep(-i)| = r$ and $0 < r < 1$, by the property
of Vandermonde matrices~\cite{MS77} we conclude that $C_i$ is
invertible for all $i=0, \ldots, p-1$ which implies that the matrices
$B_i$, $A_i$, and finally $S$ are invertible. Arguing as in the proof
of Theorem~\ref{constructionI} we have established the
informational completeness of the projectors corresponding to the
normalized vectors
$\ket{\varphi_{a,b}}$.

Next, for $r = 1-p^{-3}$ we derive the bound
\begin{equation}
\label{eq:Bound phi}
|\langle \varphi_{a,b} |
\varphi_{c, d} \rangle| \leq \frac{1+o(1)}{\sqrt{p}}, \qquad (a,b)
\ne (c,d).
\end{equation}
We have
\begin{equation}
\label{eq:Identity phi}
|\langle \varphi_{a,b} | \varphi_{c, d} \rangle|
= \sqrt{\frac{1{-}r^{2a}}{1{-}r^{2pa}}}\,
\sqrt{\frac{1{-}r^{2c}}{1{-}r^{2pc}}} \,
\left| \sum_{x\in \F_p} r^{(a+c)x} \ep(\alpha x^2 {+} \beta x) \right|,
\end{equation}
where $\alpha = a-c$
and $\beta = c-d$.

We  frequently use  that $r^t = 1 + O(t/p^3) $ for any $t = O(p^3)$.
In particular,
\begin{eqnarray*}
   \sqrt{\frac{1{-}r^{2a}}{1{-}r^{2pa}}}\,
\sqrt{\frac{1{-}r^{2c}}{1{-}r^{2pc}}}  & = & \( \sum_{ x \in \F_p}
r^{2ax} \)^{-1/2}
   \( \sum_{ x \in \F_p} r^{2cx} \)^{-1/2} \\
& \le &   \( \sum_{ x \in \F_p} r^{2px} \)^{-1}  = \(p ( 1 + O(1/p))\)^{-1}\\
& = & (1 + o(1))p^{-1}   .
\end{eqnarray*}

Furthermore,
\begin{eqnarray*}
\left| \sum_{x\in \F_p} r^{(a+c)x} \ep(\alpha x^2 {+} \beta x) \right|
&   \le &  \left| \sum_{x \in \F_p} (1 + O(1/p))\ep(\alpha x^2 {+}
\beta x) \right|
\\
&   \le &  \left| \sum_{x \in \F_p} \ep(\alpha x^2 {+} \beta x) \right|  + O(1)
\le \sqrt{p}  +
O(1).
\end{eqnarray*}
Substituting the above bounds in~\eqref{eq:Identity phi},
we derive~\eqref{eq:Bound phi}.

Next, we note have that $E_{a,b} =
\ket{\varphi_{a,b}}\bra{\varphi_{a,b}}/p$ and $G = \sum_{a,b\in \F_p}
E_{a,b}$. We now compute the frame operator $G$ and show that $G^{-1}$
is close to the identity. Similarly to the proof of
Theorem~\ref{constructionI} this implies that $F_{a,b} =
G^{-1/2}E_{a,b}G^{-1/2}$ indeed forms an
ASIC-POVM. We have
\begin{eqnarray*}
          G & = &
                  \left( \sum_{\otop{a,b\in \F_p}{x,y,\in \F_p}}
            \frac{1-r^{2a}}{p\,(1-r^{2pa})} \,
            r^{a(x+y)} \ep(a(x^2{-}y^2) + b(x{-}y)) \ket{x}\bra{y}
\right)  \\
          & = &   \left( \sum_{\otop{a\in \F_p}{x,y,\in \F_p}}
            \frac{1-r^{2a}}{1-r^{2pa}}
            r^{a(x+y)} \ep(a(x^2{-}y^2)) \delta_{x,y} \ket{x}\bra{y}
\right)   \\
          & = &
{\mathrm{diag}}\left(\sum_{a \in \F_p} \frac{1-r^{2a}}{1-r^{2pa}}
\,\,  r^{2ax}  \ : \
x \in \F_p\right).
\end{eqnarray*}

Recalling that $r = 1 - p^{-3}$,
from the Taylor expansion  we obtain that
\[
\frac{1-r^{2a}}{1-r^{2pa}} = \frac{1 + o(1)}{p}.
\]
Hence
\[
\sum_{a \in \F_p} \frac{1-r^{2a}}{1-r^{2pa}} r^{2ax} =
   \frac{1 + o(1)}{p} \sum_{a \in \F_p} r^{2a x} =  \frac{1 + o(1)}{p}
(p + O(1)) = 1 + o(1).
\]
Finally, we deduce that
\begin{eqnarray*}
p^2\tr(F_{a,b} F_{c,d})
&=&  |\langle \widetilde{\varphi}_{a,b} |
          \widetilde{\varphi}_{c,d} \rangle|^2 = |\langle \varphi_{a,b} |
          G^{-1} | \varphi_{c,d} \rangle|^2 \\
          & =  & (1 + o(1)) \left|\langle \varphi_{a,b} | \varphi_{c,d}
\rangle\right|^2
          = \frac{1+o(1)}{p},
\end{eqnarray*}
which implies that the rank one operators $F_{a,b}$
form an ASIC-POVM.
\end{proof}

\section{Relaxed MUBs and SIC-POVMs}
\label{relaxed}

\subsection{Motivation}
The constructions in the previous sections required the existence of
$n+1$ mutually unbiased bases in $\C^n$. Currently, it is not known
whether such extremal sets of mutually unbiased bases exist when $n$
is divisible by two distinct primes. We show that approximately
mutually unbiased bases exists in all dimensions. Furthermore, we show
that if we slightly relax the ASIC-POVM condition, then we can obtain
in any dimension systems of vectors that approximate SIC-POVMs.
For these constructions, we need some results about the distribution
of primes and bounds on exponential sums, which we
summarize below.

These constructions work in any dimension $n$ but attain its maximal
strength for $n$ of a certain arithmetic structure.  In particular, if
$n=p-1$ for a prime $p$, then we show the existence of AMUBs in
$\C^n$. It is not known whether maximal sets of MUBs in these
dimensions exist.

Besides the general mathematical interest in the vector systems
derived in this section, we expect that our approach and the new
technique introduced in the following will lend themselves to some
further applications in quantum information processing.

\subsection{Analytic Number Theory Background}\label{ntbackground}

First of all we recall a remarkable result of~\cite{BaHaPi} on
gaps between consecutive primes.

\begin{lemma}
\label{le:Prime Gaps} For any sufficiently large $x$, any interval
of the form $[x-x^{0.525},x]$ contains a prime number.
\end{lemma}

We now need some bounds of exponential sums with polynomials.

Let $p$ be a prime number
and let $\F_p$ be a field of $p$
elements.
We always assume that $\F_p$ is represented by the
elements $\{0,\ldots, p-1\}$.

The following statement is a variant
of the celebrated {\it Weil
bound\/},
see Example~12 of Appendix~5 of~\cite{Weil2} as well as Theorem~3
of Chapter~6 in~\cite{Li} and Theorem~5.41 and comments to
Chapter~5 of~\cite{LN}).

\begin{lemma}
\label{le:Weil} Let $\chi$ be a nontrivial multiplicative character of
$\F_p$ of order $s$
             Suppose that $G(X) \in \F_p[X]$ is not, up to a nonzero
multiplicative constant,
an $s$th power in $\F_p[X]$.
           Then for any polynomial $F(X)\in \F_p[X]$ of degree~ $d$ we have
           $$\left|\sum_{u=1}^p \ep(F(u))\chi(G(u))
                 \right|
             \leq (d + \nu -1)p^{1/2},$$
           where $\nu$ is the number of distinct roots of $G$ in the algebraic
closure of
           $\F_p$.
\end{lemma}

We now use Lemma~\ref{le:Weil} to estimate some mixed exponential
sums with two exponential functions.

\begin{lemma}
\label{le:Mixed Sum} Let $F(X) \in \F_p[X]$ be of
degree $d \geq 2$.
Then, for any integer $k$,
$$
\left|\sum_{u=1}^p \ep(F(u))  \en(ku)\right| =
\left\{\begin{array}{ll} O\(p^{2/3}\), &
\text{if}\ d =2,\\
                 O\(p^{3/4}\), &
\text{if}\ d \geq 3.\end{array}\right.
$$
\end{lemma}

\begin{proof} Because each term in our sum is of
absolute value $|\ep(F(u))  \en(ku)| = 1$,
for every integer $v \geq
0$ we have:
$$
\sum_{u=1}^p \ep(F(u))  \en(ku) = \sum_{u=1}^p \ep(F(u+v))
\en(k(u+v))  + O(v).
$$
Thus for every positive integer $m$ we have
$$
m \sum_{u=1}^p \ep(F(u))  \en(ku) = \sum_{v=0}^{m-1}\sum_{u=1}^p
\ep(F(u+v))
\en(k(u+v))  + O(m^2).
$$
Therefore
\begin{equation}
\label{eq:Sum W}
\left|\sum_{u=1}^p \ep(F(u))  \en(ku)\right|  \leq
\frac{1}{m} W + O(m),
\end{equation}
where
\begin{eqnarray*}
W & = & \left| \sum_{u=1}^p \sum_{v=0}^{m-1}
\ep(F(u+v))
\en(k(u+v)) \right| \\
& = &  \left| \sum_{u=1}^p \en(ku) \sum_{v=0}^{m-1}
\ep(F(u+v))
\en(kv) \right| \\
& \leq &  \sum_{u=1}^p  \left| \sum_{v=0}^{m-1}
\ep(F(u+v ))
\en(kv) \right|.
\end{eqnarray*}
By the Cauchy inequality we obtain
\begin{eqnarray*}
W^2 & \leq & p   \sum_{u=1}^p\left|  \sum_{v=0}^{m-1}
\ep(F(u+v)) \en(kv) \right|^2\\
& = & p \sum_{v,w=0}^{m-1}
\en(k(v-w))
\sum_{u=1}^{p} \ep(F(u+v) - F(u+w)).
\end{eqnarray*}
We now examine the polynomial $F_{v,w}(U) = F(U+v) - F(U+w)$.
By the Taylor formula
we have,
$$
F_{v,w}(U) = F(U+v) - F(U+w) = \sum_{\nu = 0}^{d-1}
\frac{F^{(\nu)}(v) -F^{(\nu)}(w)}{\nu!}  U^\nu .
$$
Clearly $F^{(d-1)}(v) = F^{(d-1)}(w)$
is possible only for $v = w$.
            For such $m$ pairs of $v$ and $w$ we estimate
the  sum over $u$ trivially as $p$.
Otherwise we estimate these
sums as $(d-2) p^{1/2}$ by Lemma~\ref{le:Weil}, getting
$$
W^2 = \left\{\begin{array}{ll} O\(mp^{2}\), &
\text{if}\ d =2,\\
O\(mp^2 + m^{2} p^{3/2}\), &
\text{if}\ d \geq 3.\end{array}\right.
$$
Thus by~\eqref{eq:Sum W}
$$
\left|\sum_{u=1}^p \ep(F(u))  \en(ku)\right|  =
\left\{\begin{array}{ll} O\( m^{-1/2} p      + m\), &
\text{if}\ d =2,\\
O\( m^{-1/2} p   +   p^{3/4}   + m\), &
\text{if}\ d \geq 3.\end{array}\right.
$$
Taking $m =\rf{p^{2/3}}$ we conclude the proof.
\end{proof}

We also need  a special case of the classical {\it Weyl bound\/}
which we present in the following form, see Lemma~3.6
of~\cite{Hua} or Lemma~2.4 of~\cite{Vau} for a similar statement
in full generality.

\begin{lemma}
\label{le:Weyl} Let $F(X) \in \F_p[X]$ be of degree $d \geq 2$.
Then for any fixed $\varepsilon > 0$ and  any integer $h \leq
p$,
$$
\left|\sum_{u=1}^h \ep(F(u)) \right|= O\(
h^{1+\varepsilon} \(\frac{1}{h} + \frac{1}{p} +
\frac{p}{h^d}\)^{1/2^{d-1}}\).
$$
\end{lemma}

\subsection{Arbitrary Dimensions}
\label{sec: any n}

We now describe a construction of a vector system,
which satisfies~\eqref{eq:delta}  exactly
and also gives a certain
approximation to~\eqref{eq:type1}.
In fact we are able to get
$n^d+1$ (rather than $n+1$) orthogonal bases
with this property, where $d\geq 1$ is any integer.

Let $\cF_d$ be the set of polynomials of the form
$$
f(X) = \sum_{\nu = 2}^{d+1} a_\nu X^{\nu}.
$$
            with integer coefficients in the range
$0 \leq a_\nu \leq n-1$, $\nu =2, \ldots, d+1$.
Thus $\# \cF_d = n^d$.
Let $p$ be the smallest prime with
$p \geq n$.
For each $f \in \cF_d$ we consider the basis
\begin{equation}
\label{eq:Basis Bf}
\cB_{f} = \{\ket{\vb_{f,1}}, \ldots, \ket{\vb_{f,n}}\}
\quad
\text{where}\quad
\ket{\vb_{f,i}} =  \frac{1}{\sqrt{n}}\(\ep\(f(u)\) \en\(iu\)\)_{u=1}^n.
\end{equation}

\begin{theorem}
\label{thm:Type1}
For any integer $d \geq 1$, the standard basis and the $n^d$ bases
$\cB_f$, $f \in \cF_d$, given by~\eqref{eq:Basis Bf} are orthonormal
and satisfy also
$$
\langle \vb_{g,j} | \vb_{f,i}  \rangle =
\left\{\begin{array}{ll} O\(n^{-1/3}\), &
\text{if}\ d=1,\\
O\(n^{-1/4}\), &
\text{if}\ d\geq 2, \end{array}\right.
$$
where $f, g \in \cF_d \cup\{0\}$, $ f\neq g$, $1 \leq i,j \leq n$.
\end{theorem}

\begin{proof} The orthonormality of each basis follows from
the identity
$$
\langle \vb_{f,j} | \vb_{f,i}  \rangle
= \frac{1}{n} \sum_{u=1}^n \en((i-j) u) = \delta_{i,j}.
$$
Clearly, if $f  \in \cF_d$ and $g=0$ (or $f=0$ and $g  \in \cF_d$)
then $|\langle \vb_{g,j} |  \vb_{f,i} \rangle| = n^{-1/2}$.
Thus it remains to estimate
$$
\langle \vb_{g,j} | \vb_{f,i} \rangle = \frac{1}{n}
\sum_{u=1}^n \ep\(f(u) - g(u)\)  \en\((i-j)u\)
$$
for  $f, g \in \cF_d$,  $f \neq g$ and $1 \leq i,j \leq n$.

Because
$$
\left|\en\(f(u) - g(u)  + (i-j)u\)\right| =1
$$
and by Lemma~\ref{le:Prime Gaps} we have
\begin{eqnarray*}
\lefteqn{
\frac{1}{n}
\sum_{u=1}^n \ep\(f(u) - g(u)\)  \en\((i-j)u\) }\\
& & \qquad= \frac{1}{n}
\sum_{u=1}^p\ep\(f(u) - g(u)\)  \en\((i-j)u\) + O(|p-n|)\\
& & \qquad=\frac{1}{n}
\sum_{u=1}^p \ep\(f(u) - g(u)\)  \en\((i-j)u\) +
O(n^{-0.475}).
\end{eqnarray*}
Hence,
$$
\langle  \vb_{g,j} |\vb_{f,i} \rangle = \frac{1}{n}
\sum_{u=1}^p \ep\(f(u) - g(u)\)  \en\((i-j)u\) +
O(n^{-0.475}).
$$
Because $f(X) - g(X)$ is a polynomial of degree at least $2$,
Lemma~\ref{le:Mixed Sum} yields
$$
\sum_{u=1}^p \ep\(f(u) -g(u)\) \en\((i-j)u\) =
\left\{ \begin{array}{ll}
O(p^{2/3})=O(n^{2/3}), & d=1,\\
O(p^{3/4}) = O(n^{3/4}), &d\geq 2,
\end{array}\right.
$$
which concludes the proof.
\end{proof}

As before, let $p$ be the smallest prime with $p \geq n$.
We consider $n^2$ vectors
\begin{equation}
\label{eq:System B}
\cB = \{\ket{\vb_{f}}\ : \ f \in \cF_2\},
\quad
\text{where}\quad
\ket{\vb_{f}} =  \frac{1}{\sqrt{n}} \(\ep\(f(u)\)\)_{u=1}^n.
\end{equation}

\begin{theorem}
\label{thm:Thype2}
Let $p$ be the smallest prime with $p \geq n$.
Then,  the vector system $\cB$ of $n^2$ vectors  given
by~\eqref{eq:System B} is normalized
and also satisfies
$$
|\langle  \vb_{g}| \vb_{f} \rangle| \leq
\(2 + O(n^{-1/10})\)n^{-1/2},
$$
where $f, g \in \cF_2$,  $f \neq g$.
\end{theorem}

\begin{proof}
Obviously, we have
$$|\langle  \vb_{f}| \vb_{f} \rangle|=1,\quad f\in \cF_2.$$
Put $h =p-n$.
Then for $f,g\in \cF_2$, $f\neq g$, we have
\begin{eqnarray*}
\lefteqn{|\langle  \vb_{g}| \vb_{f} \rangle| = \frac{1}{n}
\left|\sum_{u=1}^n \ep\( f(u)- g(u) \)\right|}\\
& & \qquad = \frac{1}{n}
\left|\sum_{u=1}^p \ep\( f(u)- g(u) \) \right|+
\frac{1}{n}
\left|\sum_{u=1}^h \ep\( f(n+u)- g(n+u) \)\right|.
\end{eqnarray*}
By Lemmas~\ref{le:Weil} and~\ref{le:Weyl} we have
\begin{eqnarray*}
|\langle  \vb_{g}| \vb_{f}\rangle| &\leq &2 p^{1/2}n^{-1} +
O\(
h^{1+\varepsilon} n^{-1} \(\frac{1}{h} + \frac{1}{p} +
\frac{p}{h^3}\)^{1/4}\) \\
& = &  2 p^{1/2}n^{-1} + O\(n^{-1}\(
h^{3/4+\varepsilon} + h^{1+\varepsilon} p^{-1/4} +
h^{1/4+\varepsilon} p^{1/4}\)\).
\end{eqnarray*}
By Lemma~\ref{le:Prime Gaps} we see that
$h = O(p^{0.525})$, therefore
$$
h^{3/4+\varepsilon} + h^{1+\varepsilon} p^{-1/4} +
h^{1/4+\varepsilon} p^{1/4}
= O(p^{2/5}) = O(n^{2/5})
$$
for sufficiently small $\varepsilon$ and sufficiently
large $p$.
Noting that
$$
p^{1/2} = \(n + O\(n^{0.525}\)\)^{1/2}  = n^{1/2} \(1 +
O\(n^{-0.475}\)\)^{1/2} =  n^{1/2} + O\(n^{0.2625}\)$$
we finish the proof.
\end{proof}

\subsection{Special Dimensions}
\label{sec:special n}

Here we give some improvements of the constructions of
Section~\ref{sec: any n} for the values of $n$ for which
the smallest prime $p$ with $p \equiv 1 \pmod n$
is sufficiently small.

Let $p$ be the smallest prime such that $p \equiv 1 \pmod n$. Let
$\cX_n$ be the set of $n$ characters of order $n$ modulo $p$ and
$\cU_n$ the subgroup of residues of order $n$ in $\F_p^\times$.  In
particular $\# \cU_n = n$.  It is know that $\cX_n$ is a cyclic group,
so for some character $\chi \in \cX_n$ all other characters of $\cX_n$
are given be the powers $\chi^i$, $i =1,\ldots n$.

For $f \in\F_p[X]$
of degree at most $d$
and the above  character $\chi \in \cX_n$ we define
\begin{equation}
\label{eq:Basis Bf special}
\cB_{f} = \{\ket{\vb_{f,1}}, \ldots, \ket{\vb_{f,n}}\}
\quad
\text{where}\quad
\ket{\vb_{f,i}} =  \frac{1}{\sqrt{n}}\(\ep(f(u)) \chi(u)^i\)_{u \in \cU_n}.
\end{equation}

Let $\cG_d$ be the set of polynomials of the form
$$
f(X) = \sum_{\nu = 1}^{d} a_\nu X^{\nu}.
$$
            with integer coefficients in the range
$0 \leq a_\nu \leq n-1$, $\nu =2, \ldots, d+1$.

\begin{theorem}
\label{thm:special n}
For any integer $d \geq 1$, the standard basis and the $n^d$ bases
$\cB_f$, $f \in \cG_d$, given by~\eqref{eq:Basis Bf special} are
orthonormal and satisfy also
$$
            |\langle  \vb_{g,j} | \vb_{f, i} \rangle|  \leq d p^{1/2}n^{-1},
$$
where $f, g \in \cF_d \cup\{0\}$, $f \neq g$,  $1 \leq i,j \leq n$.
\end{theorem}

\begin{proof}
We have,
$$
\langle  \vb_{f,j} | \vb_{f,i}\rangle =
\frac{1}{n}  \sum_{u\in \cU_n} \chi(u)^{i-j}
=\left\{\begin{array}{ll} 1, &  i = j,\\
                 0, & i \neq j,\end{array}\right. \qquad 1 \leq i,j \leq n.
$$

We also have
\begin{eqnarray*}
\langle   \vb_{g,j} | \vb_{f,i}\rangle & = &
\frac{1}{n}\sum_{u\in \cU_n} \ep(f(u)-g(u))\chi(u)^{i-j}\\
&=&
\frac{1}{p-1} \sum_{x=0}^{p-1}
\ep(f(x^{(p-1)/n})-g(x^{(p-1)/n}))  \chi(x^{(p-1)/n})^{i-j}.
\end{eqnarray*}
Now, using  Lemma~\ref{le:Weil}, we conclude the proof.
\end{proof}

\begin{corollary}\label{AMUBconstruction}
Let $p$ be a prime and let $n=p-1$. Then an AMUB exists
in dimension $n$.
\end{corollary}

\begin{proof}
         We apply Theorem~\ref{thm:special n} for $p = n+1$ and for $d=1$.
         Hence, we have $n^2+1$ orthonormal bases such that inner products
         between their components are all bounded by $(n+1)^{1/2} n^{-1} =
         n^{-1/2} +O(n^{-1})$.
\end{proof}

\section{Remarks and Open Questions}
\label{remarks}

The questions about finding SIC-POVMs and MUBs can be reformulated as
a spherical design question in the vector space $\C^n$
(see Zauner's thesis \cite{Zauner:99} and also \cite{KlRo2,RBSC}).
Thus it is possible that the techniques of~\cite{Lev}, as well as of
more recent works, see a very inspiring survey~\cite{PfZi}, may apply
to the problem of constructing systems of $n^2$ equiangular lines in
$\C^n$, that is, SIC-POVMs. In fact, it is quite possible that with
some adjustments they may also apply to MUBs.

It is widely believed, see~\cite{Wag}, but remains
unproved, that the  smallest
prime $p \equiv 1 \pmod n$ satisfies the bound
$$
p = O(n \log^2 n).
$$
In this case, the bound of Theorem~\ref{thm:special n}
becomes  $O(n^{-1/2}\log n)$.
Thus, it is quite possible that the construction of
Section~\ref{sec:special n} is always superior to those of
Section~\ref{sec: any n}.

Finally, we remark that many of the results of this paper remain
unchanged if one uses prime powers $q=p^r$ (and thus general finite
fields $\F_q$) instead of just primes $p$.  In particular,
Corollary~\ref{AMUBconstruction} holds true in this more general
setting. Hence, in summary, we have shown that ASIC-POVMs and AMUBs
exist for any prime power dimension $q$. Moreover, we have shown that
approximate versions of mutually unbiased bases and SIC-POVMs exist in
any dimension if we are slightly more liberal about our constraints on
the angles. 

\section*{Acknowledgements}
During the preparation of this paper, A.~K. was supported in part by
NSF CAREER award CCF~0347310, NSF grant CCR 0218582, a Texas A\&M TITF
initiative, and a TEES Select Young Faculty award, M.~R. was at the
Institute for Quantum Computing, University of Waterloo, where he was
supported in part by MITACS and the {\em IQC Quantum Algorithm
   Project} funded by ARO/ARDA, I.~S.\ was supported in part by ARC
grant DP0211459, A.~W.\ was supported in part by the Austrian Academy
of Sciences and by FWF grant S8313.

This work was partially done during   visits
by I.S. to the Institute for Quantum Computing (Waterloo, Canada) and
to the Johann Radon Institute for Computational
and Applied Mathematics (Linz, Austria);
the support and hospitality of these institutions are gratefully
acknowledged.

\end{document}